\pdfoutput=1
\newcommand*{\ATLASLATEXPATH}{}
 
\documentclass[cernpreprint,texlive=2016,txfonts,UKenglish,mhchem]{\ATLASLATEXPATH atlasdoc}
\usepackage[full, backend=biber, subcaption]{\ATLASLATEXPATH atlaspackage}
\usepackage{\ATLASLATEXPATH atlasbiblatex}
\usepackage{\ATLASLATEXPATH atlasmetadefs}
\usepackage[hepparticle, hepprocess, unit=false]{\ATLASLATEXPATH atlasphysics}
 
\graphicspath{{logos/}{figs/}}
 
\usepackage{TOPQ-2016-14-PAPER-defs}
 
\AtlasTitle{Measurement of the production cross-section of a single top quark
in association with a $Z$ boson in proton--proton collisions at 13~TeV with the ATLAS detector}
 
\AtlasAbstract{
The production of a top quark in association with a $Z$ boson is investigated.
The proton--proton collision data collected by the ATLAS experiment at the LHC in 2015 and 2016
at a centre-of-mass energy of $\sqrt{s} = 13\,\text{TeV}$ are used,
corresponding to an integrated luminosity of $36.1\,\text{fb}^{-1}$.
Events containing three identified leptons (electrons and/or muons) and
two jets, one of which is identified as a $b$-quark jet are selected.
The major backgrounds are diboson, $t\bar{t}$ and $Z\text{\,+\,jets}$ production.
A neural network is used to improve the background rejection and extract the signal.
The resulting significance is $4.2\sigma$ in the data and the expected significance is $5.4\sigma$.
The measured cross-section for $tZq$ production is $600 \pm 170\,\text{(stat.)} \pm 140\,\text{(syst.)}\,\text{fb}$.
}
 
\author{The ATLAS Collaboration}
 
\AtlasRefCode{TOPQ-2016-14}
 
\PreprintIdNumber{CERN-EP-2017-188}

\AtlasJournal{Phys.\ Lett.\ B.}
 
\hypersetup{pdftitle={ATLAS document},pdfauthor={The ATLAS Collaboration}}
 
\begin{document}
 
\maketitle
 
\section{Introduction}
\label{sec:intro}
At hadron colliders, the top quark is typically produced in \ttbar pairs through
the strong interaction or as a single top or antitop quark through the
electroweak interaction. The top quark was first observed via \ttbar production at the Tevatron~\cite{Abe:1995hr, Abachi:1995iq}.
This was followed by
the observation of single top-quark production~\cite{Abazov:2009ii, Aaltonen:2009jj,CDF:2014uma} in the $t$- and $s$-channels, also at the Tevatron.
The associated $\Pqt\PW$ production was first observed in \SI{8}{\TeV} proton--proton collisions at the Large Hadron Collider (LHC)~\cite{TOPQ-2012-20, CMS-TOP-12-040}.
These single-top-quark channels allow a direct determination of the
dominant $\Pqt\PW\Pqb$ vertex and of the magnitude of the
CKM matrix element \Vtb~\cite{Olive:2016xmw} using their measured cross-sections.
 
\begin{figure}[htbp]
\centering
\begin{tabular}{cc}
\raisebox{0ex}{\includegraphics[width=.33\textwidth]{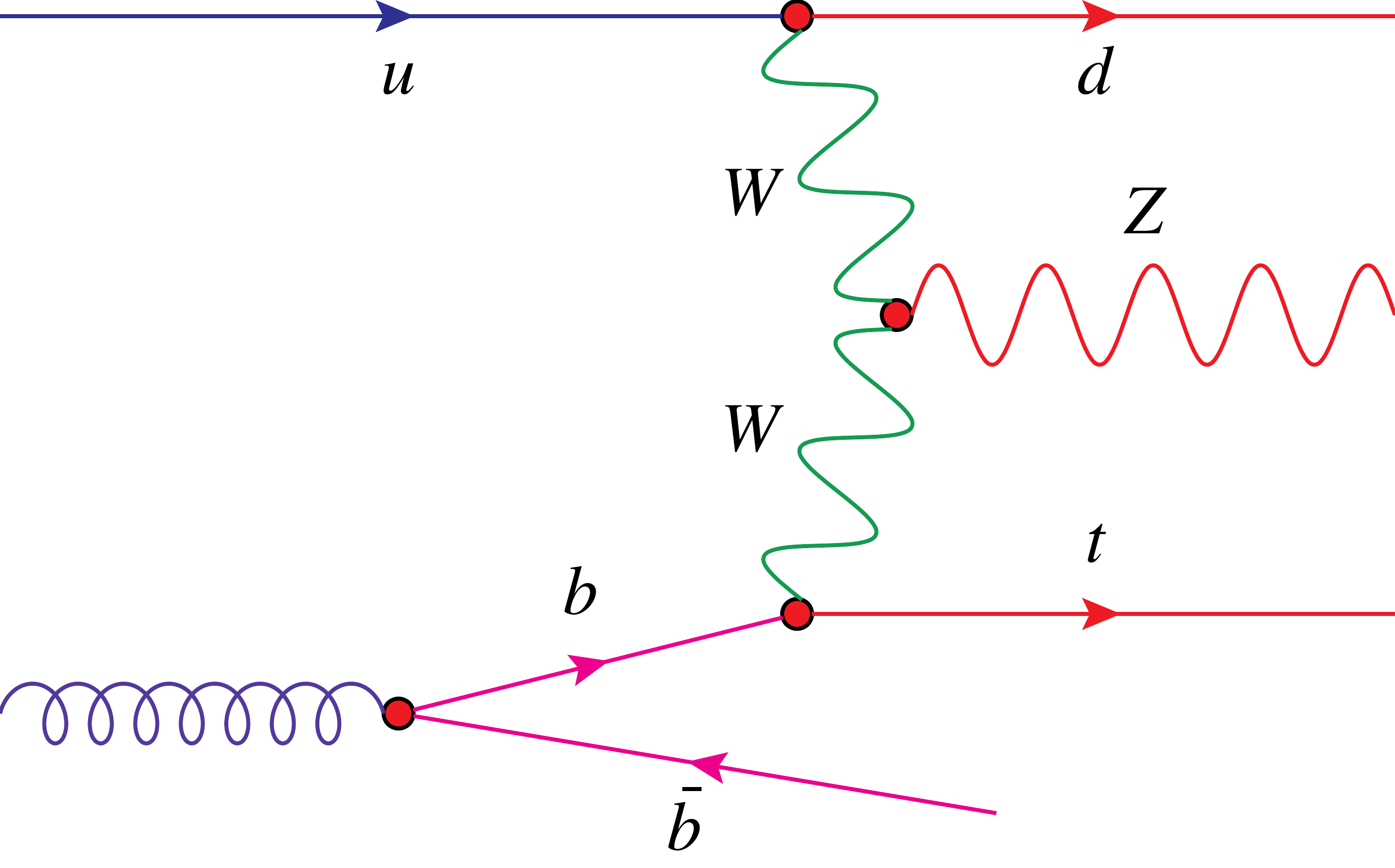}} &
\raisebox{1.2ex}{\includegraphics[width=.33\textwidth]{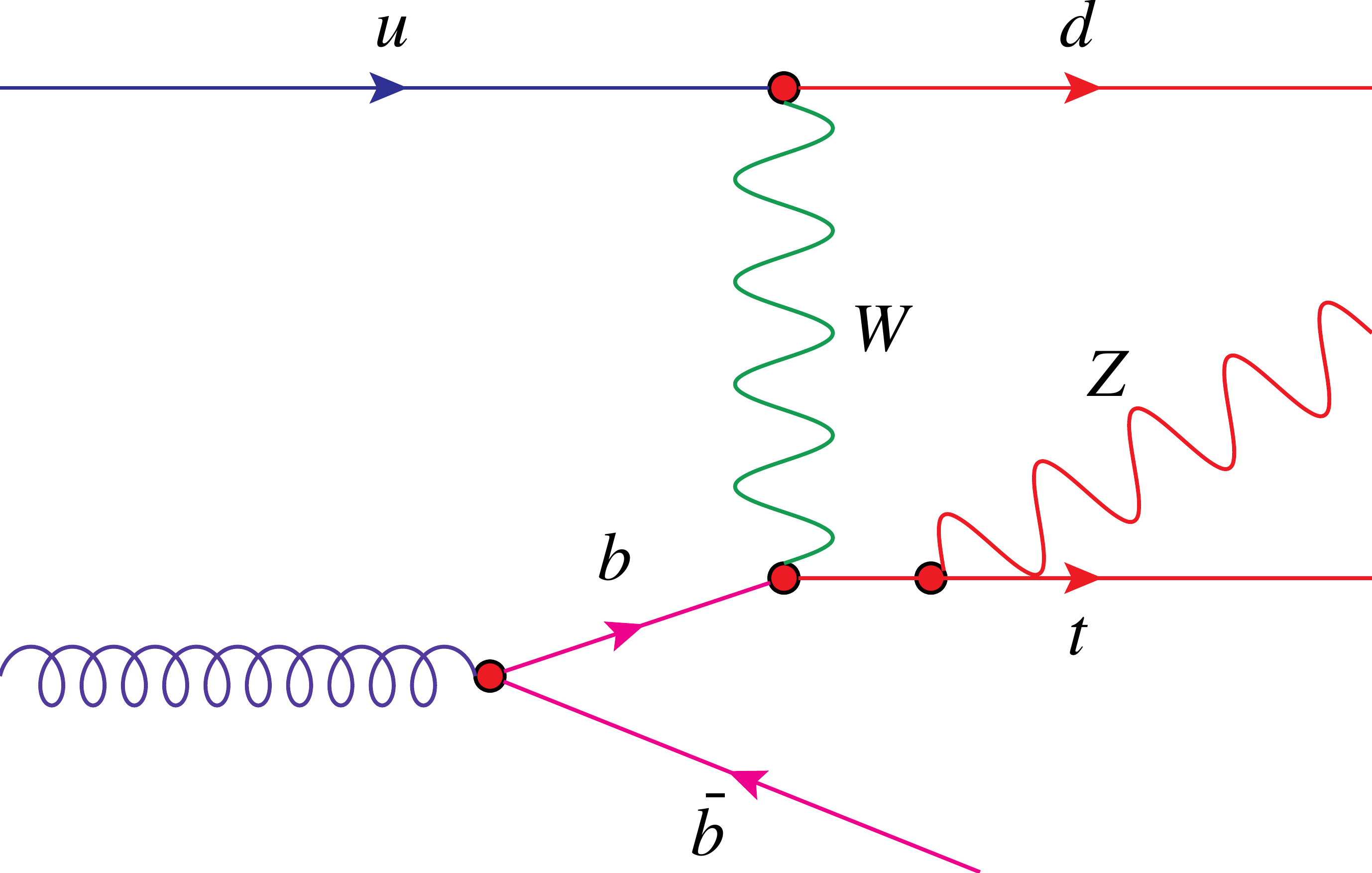}} \\
(a) & (b)
\end{tabular}
\caption{Example Feynman diagrams of the lowest-order
amplitudes for the $tZq$ process. In the four-flavour scheme, the $b$-quark originates from gluon splitting.
The largest contributing amplitude to the cross-section where the \PZ boson is coupled to the \PW boson is shown in (a) while
(b) shows one of the four diagrams with radiation off a fermion.}
\label{fig:intro:feynman}
\end{figure}

With increasing energy and integrated luminosity, the ability to study
rare Standard Model (SM) phenomena becomes possible. In the case of single top-quark
production, examples include $pp\to\Pqt\PZ\Pq$~\cite{Campbell:2013yla} and $pp\to\Pqt\PH$~\cite{Chang:2014rfa}.
The $pp\to\Pqt\PZ\Pq$ process involves \PW{}\PW{}\PZ and \Pqt{}\PZ couplings
and has not been observed so far~\cite{CMS-TOP-12-039}.
Figure~\ref{fig:intro:feynman} shows typical lowest-order Feynman diagrams for the process.
This channel probes two SM couplings in a single process,
whereas the similar final state $\ttbar\PZ$ only probes the
$\Pqt{}\PZ$ coupling. The $\ttbar\PZ$ process has been measured
by the ATLAS~\cite{TOPQ-2013-05,TOPQ-2015-22} and
CMS~\cite{CMS-TOP-14-021} collaborations.
In addition, the production of $pp\to\tZq$ is a SM background to the $\Pqt\PH$ final state~\cite{Chang:2014rfa}.
 
This Letter presents evidence of the production of a single top quark
in association with a \PZ boson in the $t$-channel process
$pp\to \Pqt{}\PZ{}\Pq$, where the \PZ boson decays into electrons or muons and the \PW boson from the top quark decays leptonically.
 
\section{ATLAS detector}
\label{sec:atlas1}
 
\newcommand{\AtlasCoordFootnote}{
ATLAS uses a right-handed coordinate system with its origin at the nominal interaction point (IP)
in the centre of the detector and the $z$-axis along the beam pipe.
The $x$-axis points from the IP to the centre of the LHC ring,
and the $y$-axis points upwards.
Cylindrical coordinates $(r,\phi)$ are used in the transverse plane,
$\phi$ being the azimuthal angle around the $z$-axis.
The pseudorapidity is defined in terms of the polar angle $\theta$ as $\eta = -\ln \tan(\theta/2)$.
Distances in the $\eta$--$\phi$ plane are measured in units of $\Delta R \equiv \sqrt{(\Delta\eta)^{2} + (\Delta\phi)^{2}}$.}
 
The ATLAS experiment~\cite{PERF-2007-01} at the LHC is a multi-purpose particle detector
with a forward-backward symmetric cylindrical geometry and a near $4\pi$ coverage in
solid angle.\footnote{\AtlasCoordFootnote}
It consists of an inner detector (ID) surrounded by a thin superconducting solenoid
providing a \SI{2}{\tesla} axial magnetic field, electromagnetic and hadron calorimeters, and a muon spectrometer.
The inner detector covers the pseudorapidity range $|\eta| < 2.5$.
It consists of silicon pixel, silicon micro-strip and transition radiation tracking detectors.
The innermost pixel layer, the insertable B-layer, was added between \RunOne and \RunTwo of the LHC, at a radius of \SI{33}{mm} around a new, thinner, beam pipe~\cite{ATL-PHYS-PUB-2015-051}.
Lead/liquid-argon (LAr) sampling calorimeters provide electromagnetic (EM) energy measurements
with high granularity.
A hadron (steel/scintillator-tile) calorimeter covers the central pseudorapidity range ($|\eta| < 1.7$).
The end-cap and forward regions are instrumented with LAr calorimeters
for both the EM and hadronic energy measurements up to $|\eta| = 4.9$.
The muon spectrometer surrounds the calorimeters and is based on
three large air-core toroidal superconducting magnets with eight coils each.
The field integral of the toroids ranges between \num{2.0} and \SI{6.0}{\tesla\metre}
across most of the detector.
The muon spectrometer includes a system of precision tracking chambers and fast detectors for triggering.
A two-level trigger system is used to select events.
The first-level trigger is implemented in hardware and uses a subset of the detector information
to reduce the accepted rate to at most \SI{100}{\kilo\hertz}.
This is followed by a software-based trigger level that
reduces the accepted event rate to \SI{1}{\kilo\hertz} on average.
 
\section{Data and simulation samples}
\label{sec:datamc}
 
The $pp$ collision data sample used in this measurement was collected with
the ATLAS detector at the LHC during the 2015 and 2016 data-taking periods,
corresponding to integrated luminosities of \SI{3.3}{\ifb} and \SI{32.8}{\ifb}, respectively, for a total of \lumi, after requiring that the detector is fully operational.
Events are considered if they were accepted by at least one of the single-muon or single-electron
triggers~\cite{ATL-DAQ-PUB-2017-001,TRIG-2016-01}.
The electron triggers select a cluster in the calorimeter matched to a track.
Electrons must then satisfy identification criteria based on a multivariate technique using a likelihood discriminant.
In 2015, electrons had to satisfy a \enquote{medium} identification requirement and have a transverse energy of $\ET > \SI{24}{\GeV}$.
In 2016, electrons had to satisfy a \enquote{tight} identification together with an isolation criterion and have $\ET > \SI{26}{\GeV}$.
To avoid efficiency loss due to isolation at high \ET, an additional trigger was used,
selecting \enquote{medium} electrons with $\ET > \SI{60}{\GeV}$.
Muons are triggered on by matching tracks reconstructed in the muon spectrometer and in the inner detector.
In 2015, muons had to satisfy a \enquote{loose} isolation requirement and have a transverse momentum of $\pT >\SI{20}{\GeV}$.
In 2016, the isolation criteria were tightened and the threshold increased to $\pT = \SI{26}{\GeV}$.
In both years, another muon trigger without any isolation requirement was used, selecting muons with $\pT > \SI{50}{\GeV}$.
 
In order to evaluate the effects of the detector resolution and acceptance on
signal and background and to estimate the SM background, a full
\GEANT{}4-based detector simulation was
used~\cite{Agostinelli:2002hh, SOFT-2010-01}.
Event generators were used to estimate the expected signal and background contributions and their uncertainties.
The top-quark mass in the event generators described below was set to \SI{172.5}{\GeV}.
Multiple inelastic \pp collisions (referred to as pile-up) are simulated with \PYTHIAV{8.186}~\cite{Sjostrand:2007gs},
and are overlaid on each Monte Carlo (MC) event.
Weights are assigned to the simulated events such that
the distribution of the number of pile-up interactions in the simulation
matches the corresponding distribution in the data.
All simulation samples are processed through the same reconstruction algorithms as the data.
 
Monte Carlo \tZq signal samples were generated at leading
order (LO) in QCD using \MGatNLOV{2.2.1}~\cite{Alwall:2014hca} in the four-flavour
scheme, treating the \Pqb-quark as massive, with the CTEQ6L1~\cite{Pumplin:2002vw}
LO parton distribution functions (PDFs). The \PZ boson was simulated to be on-shell and
off-shell \PZ/$\gamma^*$ contributions and their interference are not taken
into account.
Following the discussion in Ref.~\cite{Frederix:2012dh},
the renormalisation and factorisation scales ($\mu_\mathrm{r}$ and $\mu_\mathrm{f}$)
used in \MGatNLO are set to
$\mu_\mathrm{r}=\mu_\mathrm{f}=4 \sqrt{ m^{2}_{\Pqb}+p^{2}_{\text{T},\Pqb}}$,
where the \Pqb-quark is the external one produced from gluon splitting in the event.
This choice is motivated by the total scale dependence
being dominated by this external \Pqb-quark, shown in \Fig{\ref{fig:intro:feynman}}.
The parton shower and the hadronisation of signal events were
simulated with \PYTHIAV{6}~\cite{Sjostrand:2006za} using the Perugia2012 set of tuned parameters~\cite{Skands:2010ak}.
The \tZq total cross-section, calculated at next-to-leading order (NLO) using \MGatNLOV{2.3.3} with the NNPDF3.0\_nlo\_as\_0118~\cite{Ball2015} PDF, is
\SI{800}{\fb}, with an uncertainty of $^{+6.1}_{-7.4}\%$.
The uncertainty is computed by varying the renormalisation and factorisation scales by a factor of two and by a factor of 0.5.
 
A comparison of the event kinematics before parton showering between the LO \MGatNLOV{2.2.1} sample
and a sample generated using NLO \MGatNLOV{2.3.3} showed agreement within \SI{10}{\%},
justifying the use of a LO sample for the detector simulation.
 
Monte Carlo simulated events are used to estimate the SM background that can
produce three leptons and at least two jets in the final state. In
\ttbar production, if both \PW bosons decay into leptons (referred to as \enquote{prompt}) and either
a \Pqb- or \Pqc-hadron decays into a lepton (referred to as \enquote{non-prompt}) that is isolated, the
final state can mimic the \tZq final state. The nominal \ttbar simulated
sample was generated at NLO with the
\POWHEGBOX~\cite{Nason:2004rx, powheg, Alioli:2010xd}
event generator using the CT10 PDFs~\cite{Lai:2010vv}.
The cut-off parameter, \hdamp, for the first emission of gluons
was set to the top-quark mass.
The events were then processed using \PYTHIAV{6} to perform the fragmentation
and hadronisation, and to generate the underlying event.
 
Events from the associated production of a \ttbar pair and a boson (\PW/\PZ/\PH)
provide additional modes for the production of leptons in the final
state. For $\ttbar+\PW$ the MC simulated events were generated
using \MGatNLOV{2.2.2}~\cite{Alwall:2014hca},
while the $\ttbar+\PH$ and $\ttbar + \PZ$ MC simulated events were
generated using \MGatNLOV{2.2.3}.
The generated events were then processed
with \PYTHIAV{8}~\cite{Sjostrand:2007gs} to perform the fragmentation and hadronisation,
and to generate the underlying event,
using the NNPDF2.3LO PDF set and the A14 tune~\cite{ATL-PHYS-PUB-2014-021}.
 
Processes that include the production of \PW{}\PW, \PW{}\PZ and \PZ{}\PZ events were simulated using \SHERPAV{2.1.1} at LO with up to three additional partons and the CT10 PDF set.
In the trilepton topology, the diboson background consists mainly of \PW{}\PZ events, while the contribution to the background from \PW{}\PW final states, corresponding to the case where a jet is misidentified as a lepton, is negligible. The \PZ{}\PZ background gives a small contribution of \SI{9}{\%} of all diboson events. The gluon-induced diboson production, which amounts to about \SI{10}{\%} of the quark-induced diboson production, is therefore negligible in the \tZq signal region, and is not included in the diboson samples.
In order to estimate the systematic uncertainty,
additional diboson samples were simulated using the \POWHEGBOX generator in combination with \PYTHIAV{8} and the CTEQ6L1 PDF sets.
 
Of the aforementioned single-top-quark production channels, only the \tW channel contributes to the trilepton final state.
This sample was produced using the NLO \POWHEGBOX event generator with the CT10 PDF set.
The events were then
processed with \PYTHIAV{6} to perform the fragmentation and
hadronisation, and produce the underlying event.
A sample of \tWZ events was produced using
the \MGatNLOV{2.2.3} generator and showered with \PYTHIAV{8}, using the NNPDF3.0\_NLO PDF set and the A14 tune.
 
\section{Object reconstruction}
\label{sec:reco}
 
The reconstruction of the basic physics objects used in this analysis is described in the following.
The primary vertex is chosen as the proton--proton vertex candidate with the highest sum of the squared transverse momenta of all associated tracks with $\pT > \SI{400}{\MeV}$.
 
Electron candidates are reconstructed from energy deposits in the
electromagnetic calorimeter that match a reconstructed track~\cite{PERF-2013-03,ATL-PHYS-PUB-2015-041,
ATL-PHYS-PUB-2016-015,ATLAS-CONF-2016-024}. The
clusters are required to be within $|\eta|<2.47$ excluding
the transition region between the barrel and end-cap calorimeters at
$1.37<|\eta|<1.52$.
Electron candidates must also satisfy a
transverse energy requirement of $\et>\SI{15}{\GeV}$.
A likelihood-based discriminant is constructed from a set of
variables that enhance the electron selection, while
rejecting photon conversions and hadrons misidentified as
electrons~\cite{ATL-PHYS-PUB-2015-041}.
An $|\eta|$- and \pt-dependent
selection on the likelihood discriminant is applied, such that it
has an \SI{80}{\%} efficiency when used to identify electrons from the \PZ{}-boson decay.
This working point corresponds to an approximate rejection factor against jets
of 700 at a \pT of \SI{40}{\GeV}.
Electrons are further required to be isolated using criteria based on ID tracks and topological clusters in the calorimeter, with an isolation efficiency of \SI[parse-numbers=false]{90(99)}{\%} for  $\pT = \SI[parse-numbers=false]{25(60)}{\GeV}$.
Correction factors are applied to simulated electrons to take into account the small differences in
reconstruction, identification and isolation efficiencies between data and MC simulation.
 
Muon candidates are required to have $|\eta|<2.5$
and $\pt>\SI{15}{\GeV}$, and are reconstructed
by combining a reconstructed track from the inner detector with one
from the muon spectrometer~\cite{PERF-2015-10}.
To reject misidentified
muon candidates, primarily from pion and kaon decays, several
quality requirements are imposed on the muon candidate.
An isolation requirement based on ID tracks and topological clusters in the calorimeter is imposed, and results in an isolation efficiency of \SI[parse-numbers=false]{90(99)}{\%} for $\pT = \SI[parse-numbers=false]{25(60)}{\GeV}$.
The overall efficiency obtained for muons from \PW-boson decays in simulated $pp\to\ttbar$ events is \SI{96}{\%} and the rejection
factor for non-prompt muons with $\pt>\SI{20}{GeV}$ is approximately 600.
As for electrons, correction factors are applied to muons to account for the small differences between data and simulation.
 
Jets are reconstructed from topological clusters using the \antikt
algorithm~\cite{Cacciari:2005hq,Cacciari:2008gp} with the radius
parameter set to $R=0.4$. They are reconstructed for $\pT > \SI{30}{GeV}$ in the region with $\abseta < 4.5$. To account for inhomogeneities and the
non-compensating response of the calorimeter, the reconstructed jet energies
are corrected using \pt- and $\eta$-dependent factors that are derived
in MC simulation and validated in data.
Any remaining differences
in the jet energy scale are corrected using \emph{in situ} techniques,
where a well-defined reference object is momentum-balanced with a
jet~\cite{PERF-2012-01}.
To suppress pile-up, a discriminant called the jet-vertex-tagger (JVT) is constructed using a two-dimensional likelihood method~\cite{PERF-2014-03}.
For jets with $\pT<\SI{60}{\GeV}$ and $|\eta| < 2.4$ a JVT requirement corresponding to a \SI{92}{\%} efficiency, while rejecting \SI{98}{\%} of jets from pile-up and noise, is imposed.
 
To identify jets containing a \Pqb-hadron (\Pqb-tagging), a multivariate
algorithm
is employed~\cite{PERF-2012-04}. This algorithm uses the impact parameter
and reconstructed secondary vertex information of the tracks contained in
the jet as input for a neural network.
Due to its use of the inner detectors, the reconstruction of \Pqb-jets is done in the region with $\abseta < 2.5$.
Jets initiated by \Pqb-quarks are selected
by setting the algorithm's output threshold such that a \SI{77}{\%} \Pqb-jet
selection efficiency is achieved in simulated \ttbar events. With this setting,
the misidentification rate for jets initiated by light-flavour quarks or gluons is \SI{1}{\%}, while it is \SI{17}{\%} for jets initiated by \Pqc-quarks~\cite{ATL-PHYS-PUB-2016-012}.
Correction factors are derived and applied to correct for the small differences in \Pqb-quark selection efficiency between data and MC simulation~\cite{PERF-2012-04}.
 
The missing transverse momentum, with magnitude \met, is calculated as the negative of the vector sum of
the transverse momenta of all reconstructed objects, \pTmiss.
In addition to the identified jets, electrons and muons,
a track-based `soft' term is included in the \pTmiss calculation,
by considering tracks associated with the primary vertex in the event but not with an identified jet, electron, or muon~\cite{PERF-2014-04, ATL-PHYS-PUB-2015-027}.
 
To avoid cases where the detector response to a single physical object is reconstructed as two separate final-state objects, several steps are followed to remove such overlaps, following Ref.~\cite{TOPQ-2015-16}.
 
\section{Signal, control and validation regions}
\label{sec:evtsel}
 
The reconstructed \tZq final state consists of three charged leptons (electron and/or muon), a \Pqb-tagged jet, an additional jet and \MET.
Reconstructing the \PZ boson and the top quark is important in order to identify specific features that help to separate the signal from the background.
For example, the \PZ-boson mass distributions can contribute to the reduction of top-quark backgrounds, as these do not include a \PZ boson in the final state,
while the untagged-jet pseudorapidity distribution differs in shape between \tZq signal events and diboson and $\ttbar\PZ$ events,
which constitute some of the largest backgrounds.
 
The signal region (SR) definition reflects the \tZq final state by
selecting only events that have exactly three charged leptons, one
\btagged jet and one additional jet, referred to as the untagged jet as no \Pqb-tagging requirement is applied.
In order to better separate the \tZq signal from background, additional requirements are imposed on the properties of the selected objects. The three leptons are sorted by their \pT, irrespective of flavour, and required to have transverse momenta of at least 28, 25 and \SI{15}{\GeV}, respectively.
Both jets are required to have $\pT > \SI{30}{\GeV}$.
 
An opposite-sign, same-flavour (\OSSF) lepton pair is required in order to reconstruct the \PZ boson.
In the \mee and \emm channels, the pair is uniquely identified.
For the \eee and \mmm events, both possible combinations are considered and the pair that has the invariant mass closest to the \PZ-boson mass is chosen.
The \PW~boson
is reconstructed from the remaining lepton and the missing transverse momentum,
using as constraint the \mbox{\PW-boson} mass to evaluate the $z$ component of the neutrino momentum\footnote{
In case of an imaginary solution, the \pTmiss value is varied until one
real solution is found.
}.
The top quark is reconstructed from the reconstructed \PW boson and the \Pqb-tagged jet.
 
To suppress background sources that do not contain a \PZ boson, the invariant mass of the leptons is required to be between 81 and \SI{101}{\GeV}.
Because a \PW boson is expected in the final state,
the reconstructed transverse mass\footnote{
The transverse mass is calculated using the momentum of the lepton associated with the \PW boson, \pTmiss and the azimuthal angular difference between the two:
$m_{\text{T}}\left(\ell, \pTmiss\right) = \sqrt{2 \pT(\ell) \MET
\left[1-\cos \Delta \phi \left(\ell, \pTmiss \right) \right]}$.
}
of the \PW-boson candidate is required to satisfy $\mT(\ell,\Pgn) > \SI{20}{\GeV}$.
 
The selection criteria that define the SR are summarised in \Tab{\ref{tab:cutsummary}}.
In total, 141 events are selected using these criteria.
The criteria are modified to define validation regions,
which are used to check the modelling of the main background contributions.
Two validation regions (VR) are defined as follows:
the diboson VR uses the same event selection as the SR,
except that only one jet is required in the event and no \Pqb-tagging requirement is applied.
The \ttbar VR also uses the same selection as the SR,
except that the invariant mass of the \OSSF pair must be outside the \PZ-mass window
($\mll < \SI{81}{\GeV}$ or $\mll > \SI{101}{\GeV}$).
In addition, two control regions (CR) are defined, from which the normalisations of the diboson and the \ttbar background sources are computed, as explained in \Sect{\ref{sec:backgr}}. The diboson CR is defined in the same way as the diboson VR, except with a tighter requirement on $\mT(\ell_{\PW},\Pgn)$. The \ttbar CR instead has the same selection as the SR but it requires an opposite-sign, different-flavour (\OSDF) lepton pair and rejects events with an \OSSF pair.
 
\begin{table}[htbp]
\caption{
Overview of the requirements applied for selecting events in the signal, validation and control regions.
}
\label{tab:cutsummary}
\centering
\begin{tabular}{cccc}
\toprule
& \multicolumn{2}{c}{Common selections} & \\
\midrule
& \multicolumn{2}{c}{Exactly 3 leptons with $|\eta| < 2.5$ and $\pT> \SI{15}{\GeV}$}  &  \\
& \multicolumn{2}{c}{$\pT(\ell_1)> \SI{28}{\GeV}$, $\pT(\ell_2)> \SI{25}{\GeV}$, $\pT(\ell_3)> \SI{15}{\GeV}$} &  \\
& \multicolumn{2}{c}{$\pT(\text{jet})> \SI{30}{\GeV}$} &  \\
& \multicolumn{2}{c}{$\mT(\ell_{\PW},\Pgn) > \SI{20}{\GeV}$} &\\
\midrule
SR & Diboson VR / CR & \ttbar VR & \ttbar CR \\
\midrule
$\ge$ 1 \OSSF pair & $\ge$ 1 \OSSF pair & $\ge$ 1 \OSSF pair & $\ge$ 1 \OSDF pair \\
$|\mll - \mZ| <  \SI{10}{\GeV}$ & $|\mll - \mZ| < $ \SI{10}{\GeV} &$|\mll - \mZ| > \SI{10}{\GeV}$ & No \OSSF pair \\
2 jets, $|\eta| < $ 4.5  & 1 jet, $|\eta| < $ 4.5  & 2 jets, $|\eta| < $ 4.5  & 2 jets, $|\eta| < $ 4.5 \\
1 \Pqb-jet, $|\eta| < $ 2.5 & --- & 1 \Pqb-jet, $|\eta| < $ 2.5 & 1 \Pqb-jet, $|\eta| < $ 2.5 \\
--- & VR/CR: $\mT(\ell_{\PW},\Pgn) > 20/\SI{60}{\GeV}$ & --- & --- \\
\bottomrule
\end{tabular}
\end{table}
 
\section{Background estimation}
\label{sec:backgr}
 
Different SM processes are considered as background sources for this analysis. These are either processes such as diboson or \ttVH production,
in which three or more prompt leptons are produced,
or processes with only two prompt leptons in the final state (such as \Zjets and \ttbar production)
and one additional non-prompt or \enquote{fake} lepton that meets the selection criteria.
Such non-prompt or fake leptons can originate from decays of bottom or charm hadrons,
a jet that is misidentified as an electron, leptons from kaon or pion decays,
or electrons from photon conversions.
 
The dominant source of background originates from diboson production.
This consists mainly of \PW{}\PZ events with a small fraction of \PZ{}\PZ events in which the fourth lepton is missed
(roughly \SI{9}{\%} of the total number of diboson events).
Studies in the diboson VR indicated that the number of events predicted by the \SHERPA MC samples is lower than the number observed.
The kinematic distributions are otherwise well described.
Hence, the total number of diboson events predicted by the \SHERPA samples is scaled by a factor of 1.47,
leading to an expected number of diboson events in the SR of 53.
The scale factor is derived from the diboson CR, defined in \Sect{\ref{sec:evtsel}}, by computing the data-to-MC ratio for events that satisfy the condition $\mTW > \SI{60}{\GeV}$.
This selection is applied in order to reduce the \Zjets contamination and ensure a diboson-dominated region.
The uncertainty in the scale factor is estimated by varying the requirement for the \mTW selection.
An additional uncertainty in the diboson estimate is assigned by evaluating the difference in the number of events in the signal region when using the default \SHERPA samples and a set of \POWHEG samples.
This results in an estimated uncertainty of \SI{30}{\%}, also taking into account the extrapolation of the scale factor from the CR to the SR.
 
The main sources of non-prompt or fake-lepton background for this analysis are \ttbar and \Zjets events.
These two contributions are evaluated separately.
This choice is motivated by MC generator-level studies showing that although very similar in origin, the source of the non-prompt or fake lepton is usually different for processes involving top quarks compared to \Zjets events.
For \ttbar events, in most cases, it is the softer of the two leptons assigned to the reconstructed \PZ boson,
while for \Zjets events it is the lepton not assigned to the \PZ boson.
 
In order to take into account a possible difference between data and MC simulation for \ttbar events,
the number of events containing a non-prompt or fake lepton in the MC simulation
is scaled by a data/MC factor that is derived in the \ttbar CR defined in \Sect{\ref{sec:evtsel}}.
This \ttbar control region and the signal region have very similar non-prompt lepton compositions.
Requiring a pair of opposite-sign, different-flavour leptons,
and rejecting events with an \OSSF pair, ensures that there is no contamination from \Zjets events and from the SR.
Different electron--muon invariant mass windows around the \PZ mass, with widths ranging from \SI{20} {\GeV} to \SI{60}{GeV}, were investigated and the average of the obtained factors is used for scaling the \ttbar background in the signal region.
The total uncertainty in the scaling factor is calculated taking into account this variation
and the statistical uncertainty of the sample.
This leads to a data/MC scale factor of $1.21 \pm  0.51$.
Deriving separate factors depending on the fake lepton's flavour or on the lepton \pt was also investigated.
All approaches are consistent with each other within the assigned uncertainties.  The expected number of \ttbar events in the SR is \SI{18 \pm 9}{}. According to the MC prediction, the \tW contribution is found to be less than one event.
 
A data-driven technique called the fake-factor method is used to estimate the \Zjets background contribution.
A region defined by selecting events with $\mTW < \SI{20}{\GeV}$ is used for deriving the fake factors.
Since it is observed that the number of non-prompt or fake electrons and muons can be very different,
the estimation is done separately for the electron and muon channel.
Fake factors are defined as the ratio of data events that have three isolated leptons to events in which one of the leptons fails the isolation requirement.
They are derived in bins of the \pT of the lepton not associated with the \PZ boson.
According to MC simulation, this lepton is in over \SI{95}{\%} of the cases the non-prompt or fake lepton.
These factors are then applied to events passing the signal region selection (including a $\mTW > \SI{20}{\GeV}$ cut) that have one of the three leptons failing the isolation requirement.
Contamination from other background sources, which is about \SI{50}{\%} and mainly coming from \ttbar, is taken into account and subtracted before making the final \Zjets estimate.
The expected number of \Zjets events in the SR is 37.
Different sources of uncertainty are investigated, including consistency checks of the fake-factor method using MC \Zjets samples,
the effect of changing the diboson scale factor and the statistical uncertainties in the estimated and observed number of events.
All these amount to a total uncertainty of \SI{40}{\%}.
 
The expected \ttV, \ttH and \tWZ contributions are evaluated from the MC samples normalised to their predicted NLO cross-sections~\cite{Alwall:2014hca}.
The \ttVH contribution is approximately \SI{10}{\%} of the total background estimate, while \tWZ events amount to \SI{3}{\%}. The expected number of \ttWZH events is \SI{20 \pm 3}{}.
The uncertainty in the predictions is taken to be \SI{13}{\%}~\cite{Alwall:2014hca}.
 
\section{Multivariate analysis}
\label{sec:nn}
 
A multivariate analysis is used to separate the signal from the large number of background events.
The neural-network package NeuroBayes~\cite{feindt-2004,Feindt:2006pm} is used,
which combines a three-layer feed-forward neural network with a complex robust preprocessing.
Several variables are combined into one discriminant,
then mapped onto the interval $[0,1]$, such that background-like events have an output value, \ONN,
closer to 0 and signal-like events have an output closer to 1.
All background processes are considered in the training except \ttbar production,
due to the very small number of available MC events that meet the selection criteria.
Only variables that provide separation power and are well modelled are taken into account in the final neural network (NN).
For the NN training, the ten variables with the highest separation power are used.
These variables are explained in the order of their importance in \Tab{\ref{tab:trainingVars}}.
They include simple variables, such as the \pT and $\eta$ of jets and of the lepton not associated with the \PZ boson.
Information about the reconstructed \PW boson, \PZ boson and top quark,
such as their \pT as well as their masses, is also used.
In addition, the $\Delta R$ between the untagged jet and the \PZ boson is employed as an input.

\begin{table}[htbp]
\caption{Variables used as input to the neural network, ordered by their separation power.
}
\label{tab:trainingVars}
\centering
\begin{tabular}{ll}
\toprule
Variable
& Definition\\
\midrule
$|\eta(\text{j})|$
& Absolute value of untagged jet $\eta$ \\
$\pT(\text{j})$
& Untagged jet \pT \\
$m_{\Pqt}$
& Reconstructed top-quark mass\\
$\pT(\Pl^{\PW})$
&  \pT of the lepton from the \PW-boson decay\\
$\Delta R(\text{j},\PZ)$
& $\Delta R$ between the untagged jet and the \PZ boson\\
$m_{\text{T}}(\Pl,\MET)$
&Transverse mass of \PW boson \\
$\pT(\Pqt)$
& Reconstructed top-quark \pT \\
$\pT(\Pqb)$
& Tagged jet \pT\\
$\pT(\PZ)$
& \pT of the reconstructed \PZ boson\\
$|\eta(\Pl^{\PW})|$
& Absolute value of $\eta$ of the lepton coming from the \PW-boson decay \\
\bottomrule
\end{tabular}
\end{table}
 
\begin{figure}[htbp]
\centering
\begin{subfigure}[b]{.49\textwidth}
\includegraphics[width=\textwidth]{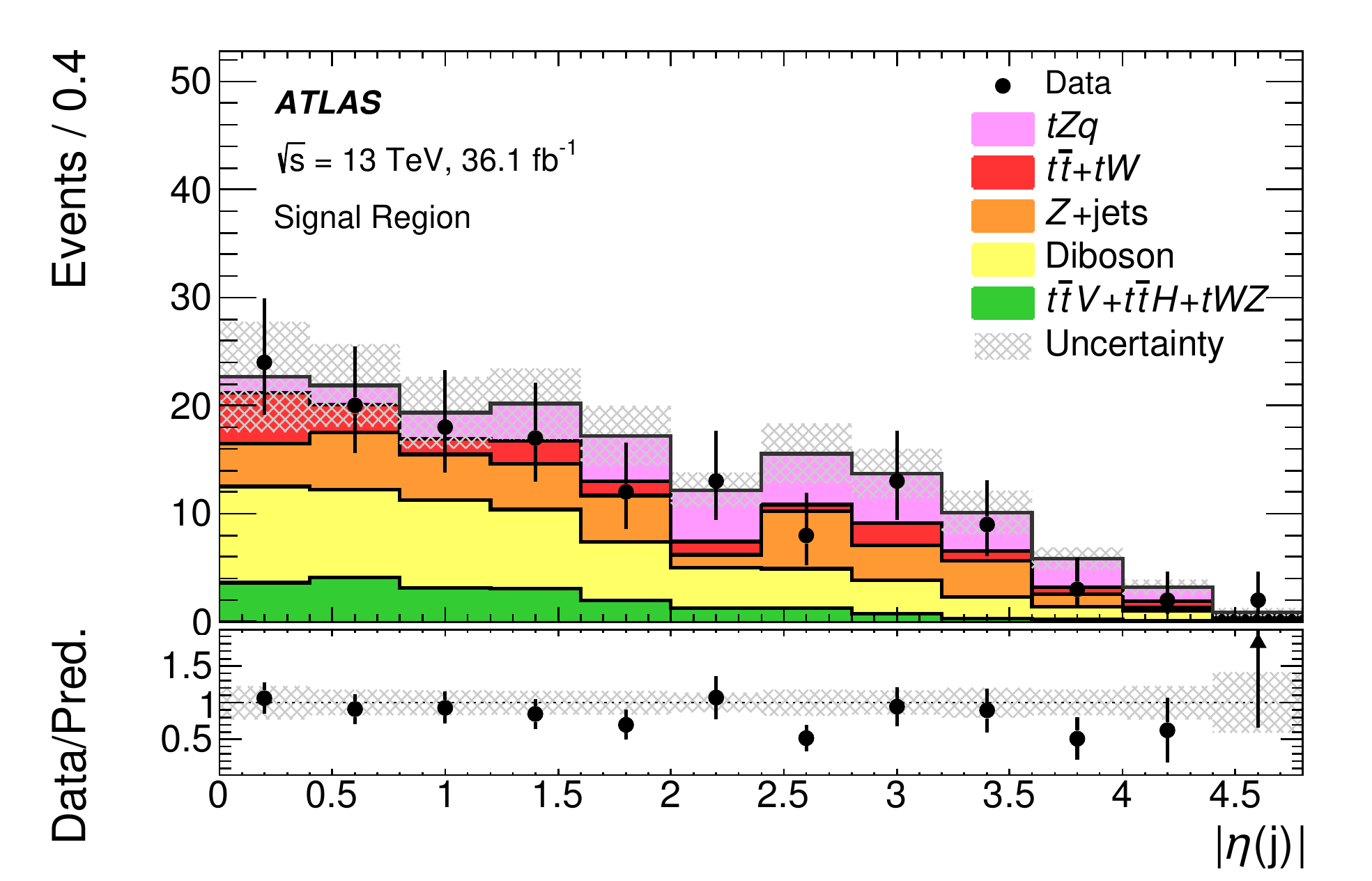}
\end{subfigure}
\begin{subfigure}[b]{.49\textwidth}
\includegraphics[width=\textwidth]{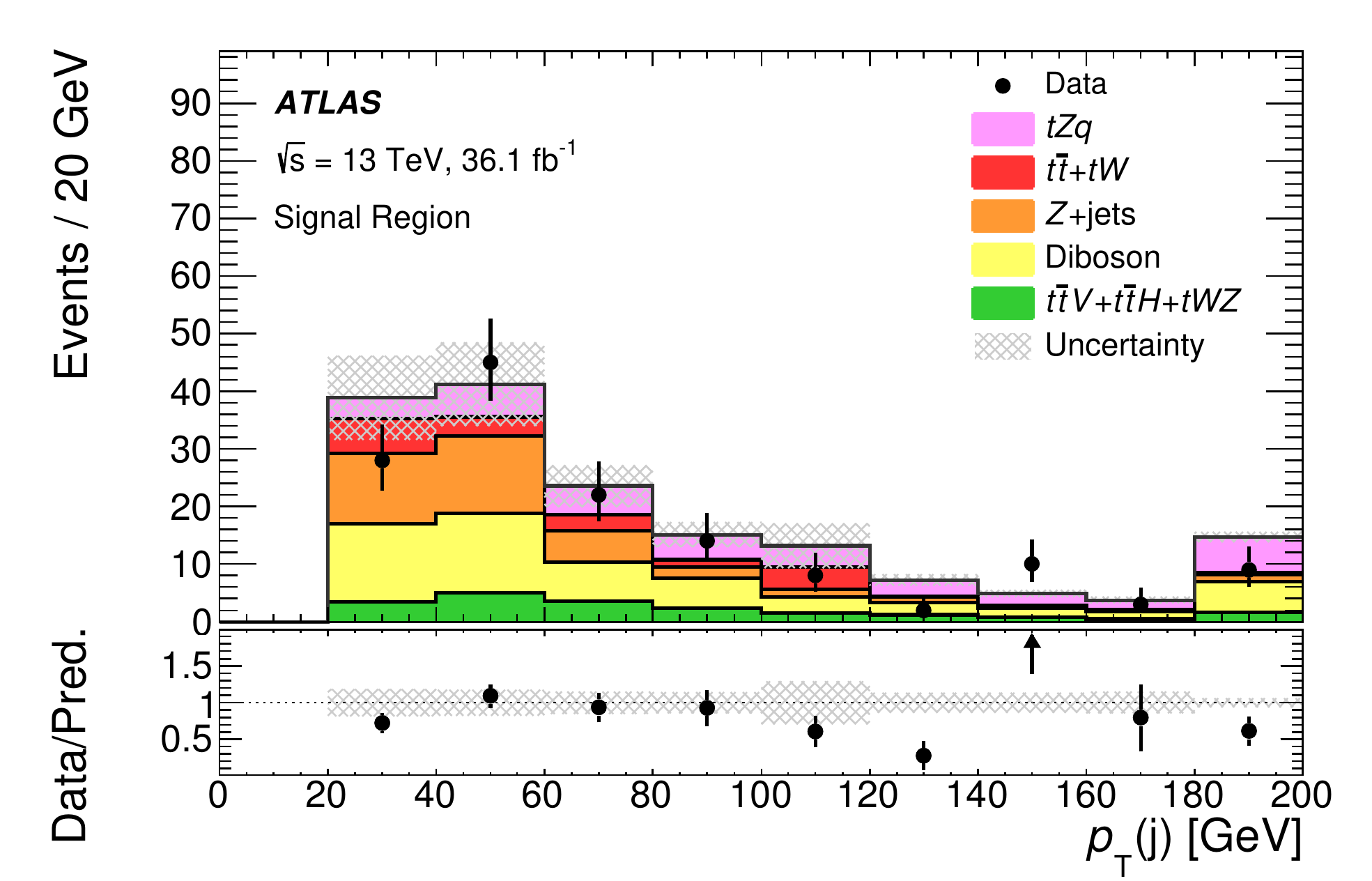}
\end{subfigure}
\vskip\baselineskip
\begin{subfigure}[b]{.49\textwidth}
\includegraphics[width=\textwidth]{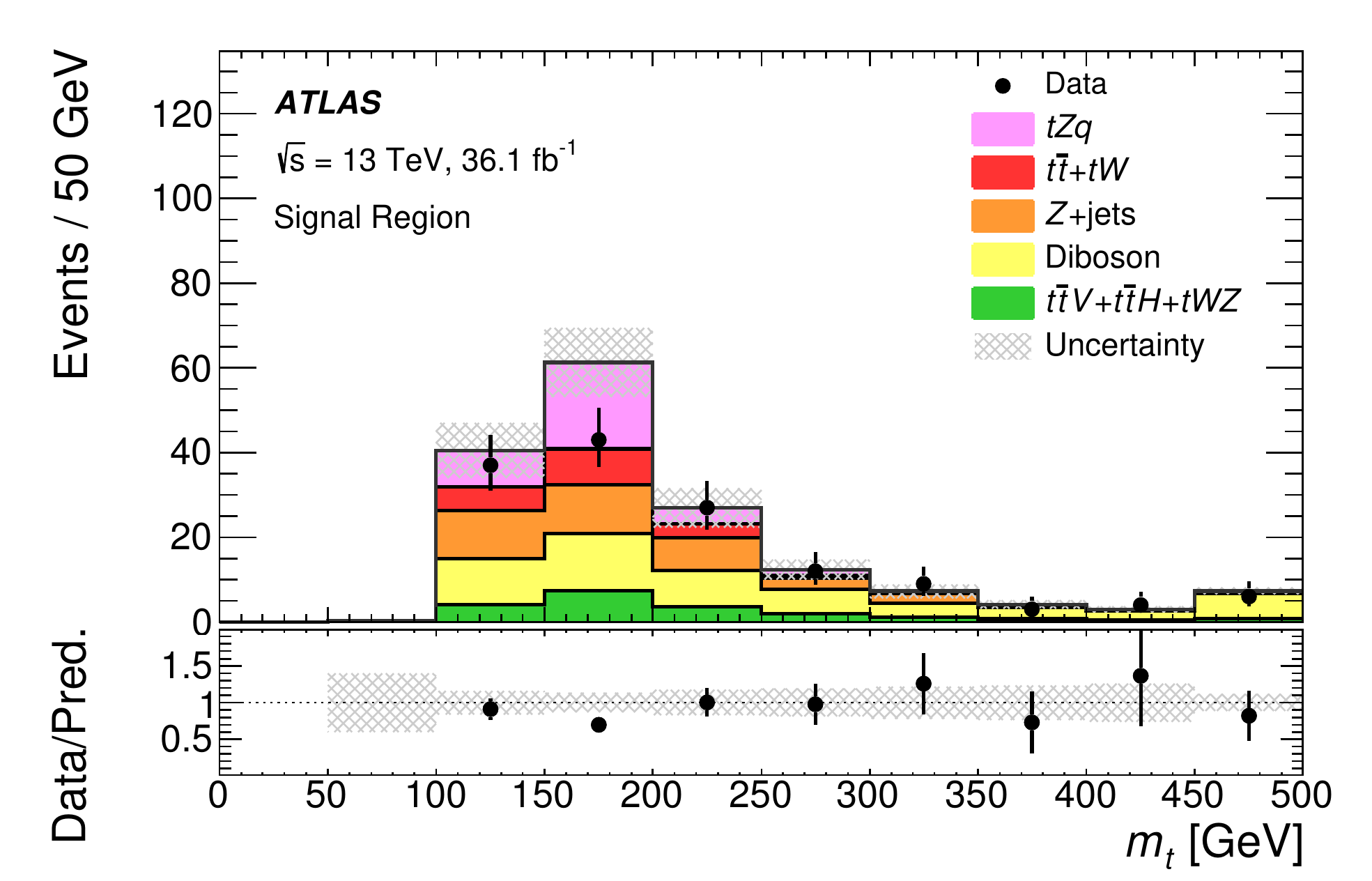}
\end{subfigure}
\begin{subfigure}[b]{.49\textwidth}
\includegraphics[width=\textwidth]{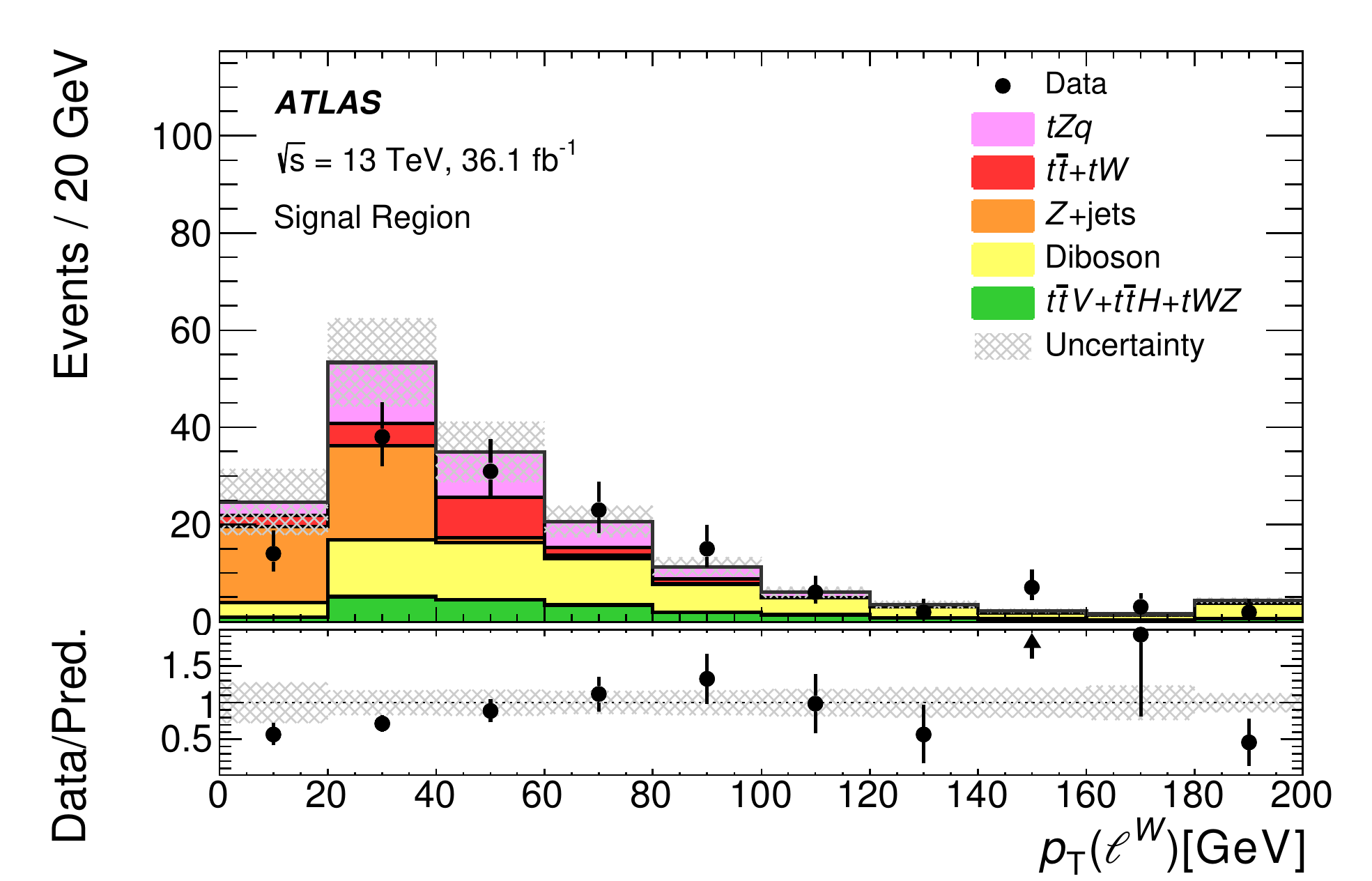}
\end{subfigure}
\caption{Comparison of the data and the signal+background model for the neural-network training variables with the highest separation power.
Signal and backgrounds are normalised to the expected number of events. The $Z\text{\,+\,jets}$ background is estimated using a data-driven technique.
The uncertainty band includes the statistical uncertainty and the uncertainties in the backgrounds derived in Section \ref{sec:backgr}.
The rightmost bin includes overflow events.
}
\label{fig:plot_NNinputs_SR}
\end{figure}

The modelling of the input variables is checked both in the validation regions defined in \Tab{\ref{tab:cutsummary}} and in the signal region.
The distributions of some input variables in the signal region are shown in \Fig{\ref{fig:plot_NNinputs_SR}}, normalised to the expected number of events,
including the scale factors determined in \Sect{\ref{sec:backgr}}.
Good agreement between data and the prediction is observed.
 
\begin{figure}[htbp]
\centering
\begin{subfigure}[b]{.49\textwidth}
\includegraphics[width=\textwidth]{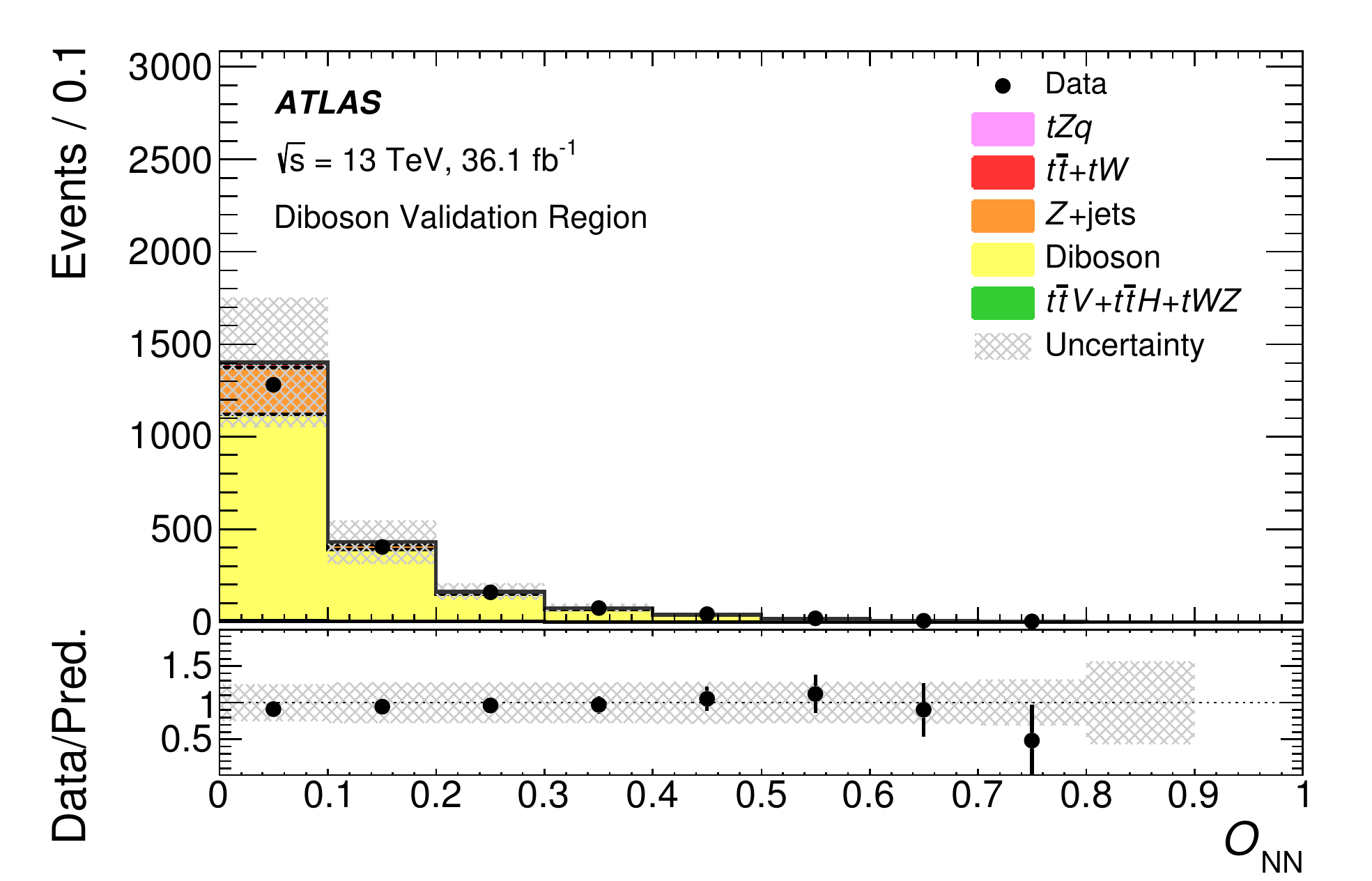}
\end{subfigure}
\begin{subfigure}[b]{.49\textwidth}
\includegraphics[width=\textwidth]{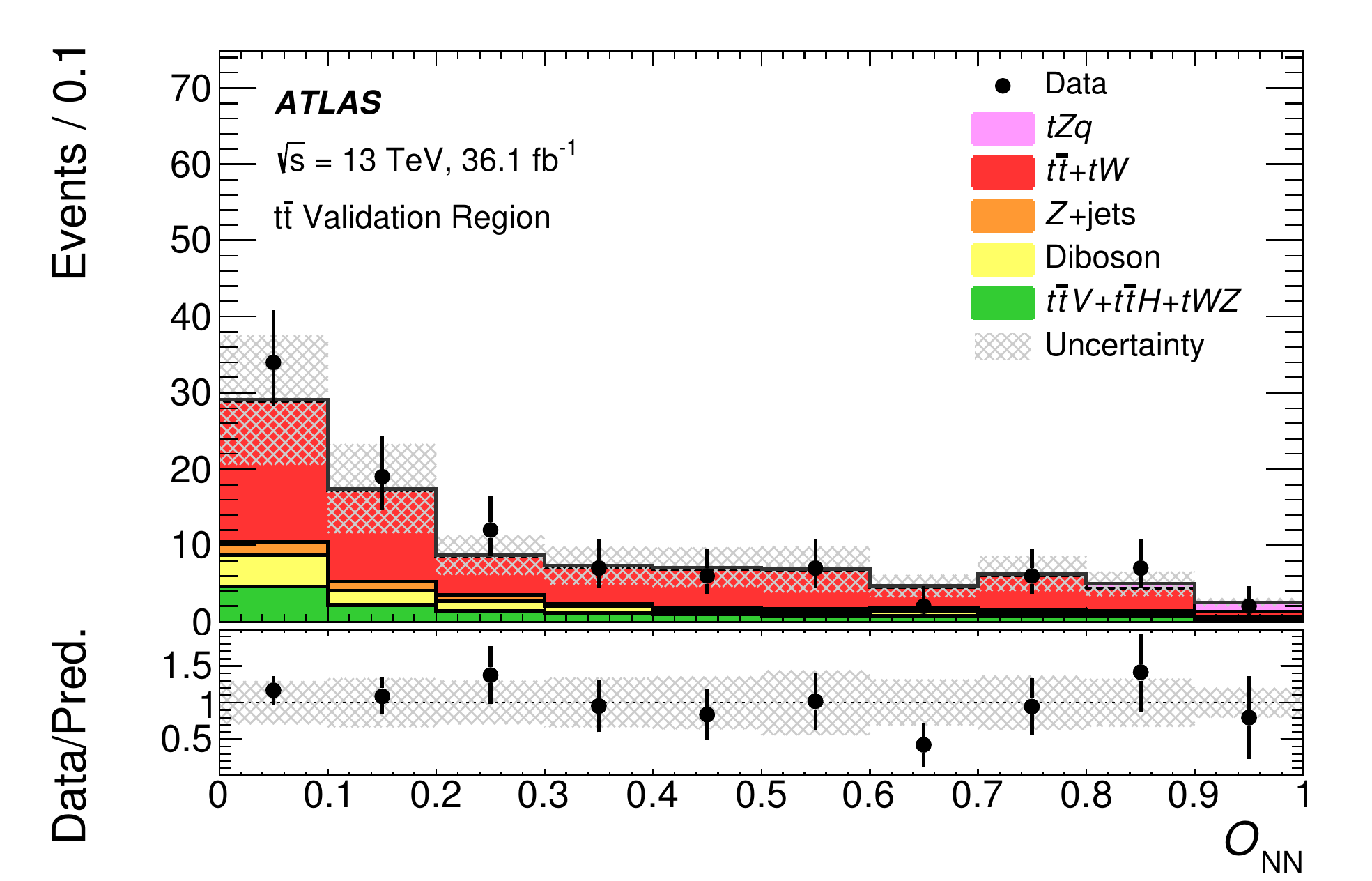}
\end{subfigure}
\caption{Neural-network output distribution of the events in the diboson (left) and $t\bar{t}$ (right) validation regions.
Signal and backgrounds are normalised to the expected number of events. The $Z\text{\,+\,jets}$ background is estimated using a data-driven technique.
The uncertainty band includes the statistical uncertainty and the uncertainties in the backgrounds derived in Section \ref{sec:backgr}.
}
\label{fig:plot_NNoutput_VR}
\end{figure}

The output of the NN is checked in the validation regions, shown in \Fig{\ref{fig:plot_NNoutput_VR}}.
Good agreement between the expected and observed numbers of events
and in the shape of the NN output distribution are seen, demonstrating reliable background modelling.
NeuroBayes includes extensive protection against overtraining and several further checks confirm that it functions well.
 
\section{Systematic uncertainties}
\label{sec:sys}
 
Systematic uncertainties in the normalisation of the individual backgrounds
and in the signal acceptance,
as well as uncertainties in the shape of the NN distributions,
are taken into account when determining the \tZq cross-section.
For uncertainties where variations as a function of the NN distribution are consistent
with being due to statistical fluctuations,
only the normalisation difference is taken into account.
The uncertainties are split into the following categories:
 
\paragraph{Reconstruction efficiency and calibration uncertainties} Systematic uncertainties affecting the reconstruction and energy calibration of jets, electrons and muons are propagated through the analysis.
The dominant sources of uncertainty for this measurement are the jet energy scale (JES) calibration,
including the modelling of pile-up, and the \Pqb-jet tagging efficiencies.
 
The uncertainties due to lepton reconstruction, identification, isolation requirements and trigger efficiencies are estimated using tag-and-probe methods in \Zll events.
Correction factors are derived to match the simulation to observed distributions in collision data and associated uncertainties are estimated.
Uncertainties in the lepton momentum scale and resolution are also assessed using \Zll events~\cite{PERF-2013-05,ATL-PHYS-PUB-2015-041,PERF-2015-10}.
 
Several components of the JES uncertainty are considered~\cite{PERF-2012-01,ATL-PHYS-PUB-2015-015}.
Uncertainties derived from different dijet-\pT-balance measurements as well as uncertainties associated with other in situ calibration techniques are considered.
Furthermore, the presence of nearby jets and the modelling of pile-up affect the jet calibration.
The uncertainty in the flavour composition covers effects due to the difference in quark–gluon composition between the jets used in the calibration and the jets used in this analysis.
Also an uncertainty due to the different calorimeter responses to light-quark and gluon jets is taken into account.
Finally, the JES uncertainty is estimated for \Pqb-jets by varying the modelling of \Pqb-quark fragmentation.
The uncertainty in the jet energy resolution (JER) and the one associated with the  JVT requirement are also considered~\cite{PERF-2011-04}.
The jet-related uncertainties with the highest impact on the final result are the JER and the flavour composition.
 
The impact of a possible miscalibration on the soft-track component of \MET is derived from data–MC comparisons of the \pT balance between the hard and soft \MET components~\cite{ATL-PHYS-PUB-2015-027}. The uncertainty associated with the leptons and jets is propagated from the corresponding uncertainties in the energy/momentum scales and resolutions, and it is classified together with the uncertainty associated with the corresponding objects.
 
Since the analysis makes use of \Pqb-tagging, the uncertainties in the \Pqb-tagging efficiency and the mistag rate are taken into account. These uncertainties were determined using $\sqrt{s}$ = 8 TeV data as described in Ref.~\cite{ATLAS-CONF-2014-004} for \Pqb-jets and Ref.~\cite{ATLAS-CONF-2014-046} for light jets, with additional uncertainties to account for the presence of the newly added inner layer of the pixel detector and the extrapolation to $\sqrt{s} = \SI{13}{\TeV}$.
 
\paragraph{Signal PDF and radiation}
The systematic effects due to uncertainties in the parton distribution functions are taken into account for the signal. As it was generated at LO, the uncertainty is evaluated using the 30 eigenvectors of the NNPDF3.0\_lo\_as\_0118~\cite{Ball2015} PDF set, in the four-flavour scheme. The events are reweighted according to each of the PDF uncertainty eigenvectors.
As a cross-check, the PDF uncertainty is also evaluated following the updated PDF4LHC recommendation~\cite{Butterworth:2015oua} by using the PDF4LHC15 NLO PDF set. This has a smaller effect; hence the uncertainty from the LO PDF set is used.
 
Variations of the amount of additional radiation are studied by changing the hard-scatter scales and the
scales in the parton shower simultaneously in the \tZq sample.
A variation of the factorisation and renormalisation
scale by a factor of two is combined with the
Perugia2012 set of tuned parameters with lower radiation (P2012radLo) than the
nominal set; while a variation of both
scales by a factor of 0.5 is combined with the
Perugia2012 set of tuned parameters with higher radiation (P2012radHi).
 
\paragraph{Luminosity}
The uncertainty in the combined 2015+2016 integrated luminosity is \SI{2.1}{\%}.
It is derived, following a methodology similar to that detailed in Ref.~\cite{DAPR-2013-01}, from a calibration of the luminosity scale using $x$--$y$ beam-separation scans performed in August 2015 and May 2016.
 
The effects of the above uncertainties on the number of signal events are summarised in \Tab{\ref{tab:fit_data_tZsysTable}}. This does not include the impact of the background uncertainties.
 
\paragraph{Background} The uncertainties in the normalisation of the various background processes use the uncertainty estimated in Section~\ref{sec:backgr}.
For the \ttbar sample, the systematic effects due to uncertainties in the scale and the amount of radiation are included.
 
\begin{table}[htbp]
\caption{Breakdown of the impact of the systematic uncertainties on the number of $tZq$ signal events
in order of decreasing effect.
Details of the systematic uncertainties are provided in the text.
MC statistics refers to the effect of the limited size of the MC samples used.
}
\label{tab:fit_data_tZsysTable}
\centering
\sisetup{round-mode=places, round-precision=1}
\begin{tabular}{lS}
\toprule
Source & \multicolumn{1}{c}{Uncertainty [\%]} \\
\midrule
\tZq radiation	&	\pm 10.8	\\
Jets	&	\pm 4.61	\\
$b$-tagging	&	\pm 2.87	\\
MC statistics	&	\pm 2.76	\\
\tZq PDF	&	\pm 2.2	\\
Luminosity	&	\pm 2.1	\\
Leptons	&	\pm 2.09	\\
\MET	&	\pm 0.26	\\
\bottomrule
\end{tabular}
\end{table}
 
\section{Results}
\label{sec:results}
 
Using the 141 selected events,
a maximum-likelihood fit is performed to extract the \tZq signal strength, $\mu$, defined as the ratio of the measured signal yield to the NLO Standard Model prediction.
The statistical analysis of the data employs a binned likelihood function $\mathcal{L} (\mu,\vec{\theta})$,
constructed as the product of Poisson probability terms, to estimate $\mu$~\cite{Cowan:2010js}.
The likelihood is maximised on the NN output distribution in the signal region.
The background normalisations are allowed to vary within the uncertainties given in Section~\ref{sec:backgr}.
 
The impact of systematic uncertainties on the expected numbers of signal and background events is described by nuisance parameters, $\vec{\theta}$,
which are each parameterised by a Gaussian or log-normal constraint for each bin of the NN output distribution.
If the variation of the uncertainty in each bin is consistent with being due to statistical fluctuations,
only the overall change in normalisation is included as a nuisance parameter.
The uncertainties are set to be symmetric in the fit,
using the average of the variations up and down.
The expected numbers of signal and background events in each bin are functions of $\vec{\theta}$.
The test statistic, $q_\mu$, is constructed according to the profile likelihood ratio:
$q_\mu = - 2 \ln [\mathcal{L} (\mu, \hat{\hat{\vec{\theta}}})/\mathcal{L} (\hat{\mu}, \hat{\vec{\theta}})]$,
where $\hat{\mu}$ and $\hat{\vec{\theta}}$ are the parameters that maximise the likelihood,
and $\hat{\hat{\vec{\theta}}}$ are the nuisance parameter values that maximise the likelihood for a given $\mu$.
This test statistic is used to determine a probability for accepting the background-only hypothesis for the observed data.
 
\begin{figure}[htbp]
\centering
\includegraphics[width=0.75\textwidth]{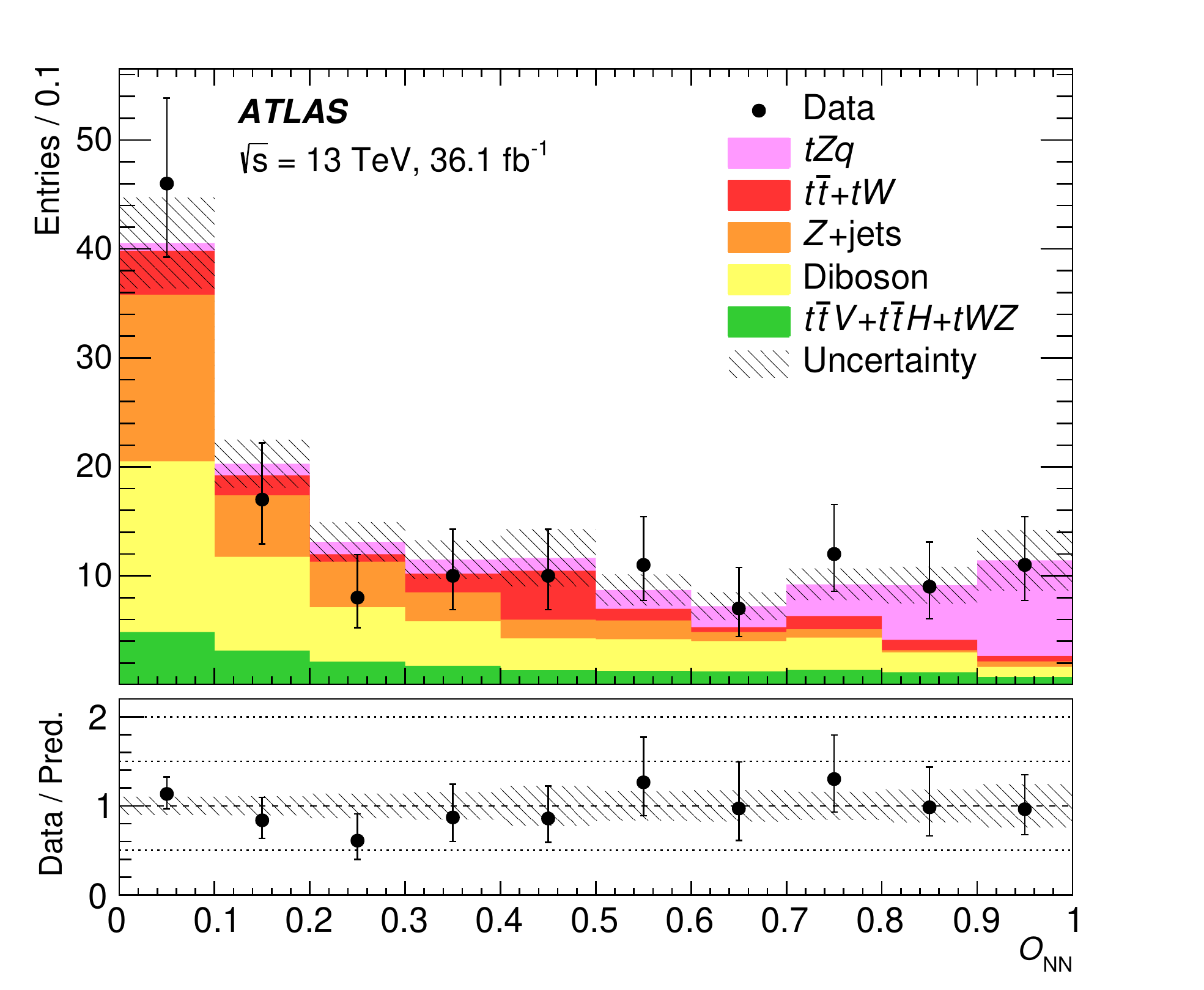}
\caption{Post-fit neural-network output distributions in the signal region. Signal and backgrounds are normalised to the expected number of events after the fit. The uncertainty band includes both the statistical and systematic uncertainties as obtained by the fit.
}
\label{fig:plot_NNoutput_SR_postfit}
\end{figure}

Figure~\ref{fig:plot_NNoutput_SR_postfit} shows the NN discriminant in the signal region with background normalisations, signal normalisation and nuisance parameters adjusted
by the profile likelihood fit.
 
The results for the numbers of fitted signal and background events are summarised in \Tab{\ref{tab:fit_data_yieldsTable}}.
The table also shows the result of a fit to the Asimov dataset~\cite{Cowan:2010js}.
The total uncertainty in the number of fitted events includes the effect of correlations, which are large among the background sources, as the \ONN distributions have a similar shape.
The strongest correlation is found to be between the diboson and the \Zjets contributions and it is about \SI{-0.5}{} for both the Asimov dataset and the data.
 
\begin{table}[htbp]
\caption{Fitted yields in the signal region for the Asimov dataset and the data.
The fitted numbers of events contain the statistical plus systematic uncertainties.
}
\label{tab:fit_data_yieldsTable}
\centering
\sisetup{round-mode=places, round-precision=0}
\begin{tabular}{l *{2}{@{\hspace*{2em}}S[table-format=3.0]@{$\,\pm\,$}S[table-format=2.0]@{\hspace*{2em}}}}
\toprule
{Channel}           &  \multicolumn{4}{c}{Number of events}\\
& \multicolumn{2}{@{\hspace*{-3em}}c}{Asimov dataset} & \multicolumn{2}{@{\hspace*{-1em}}c}{Data}\\
\midrule
\tZq            & 35.21 & 9.19 & 25.75 & 8.29 \\
\ttbar + \tW    & 17.95 & 6.90 & 16.92 & 6.80 \\
\Zjets          & 36.89 & 11.36 & 33.53 & 10.85 \\
Diboson         & 52.68 & 12.95 & 47.68 & 12.48 \\
\ttWZH          & 19.88 & 3.17 & 18.93 & 3.06 \\
\midrule
Total           & 162.61 & 11.82 & 142.81 & 10.71 \\
\bottomrule
\end{tabular}
\end{table}
 
After performing the binned maximum-likelihood fit and estimating the total uncertainty,
the fitted value for $\mu$ is \num[parse-numbers=false]{0.75 \pm 0.21 \stat \pm 0.17 \syst \pm 0.05\;\text{(th.)}}.
The quoted theory (th.) uncertainty in $\mu$ includes the \tZq NLO cross-section uncertainty given in \Sect{\ref{sec:datamc}}.
This is not taken into account when evaluating the cross-section.
The statistical uncertainty in the cross-section is determined by performing a fit to the data,
including only the statistical uncertainties.
The total systematic uncertainty is determined by subtracting this value in quadrature from the total uncertainty.
The cross-section for \tZq production is measured to be \tZxsect, assuming a top-quark mass of $m_{\Pqt}$ = \SI{172.5}{\GeV}.
 
The probability $p_0$ of obtaining a result at least as signal-like as observed in the data if no signal were present is calculated using the test statistic $q_{\mu = 0}$ in the asymptotic approximation~\cite{Cowan:2010js}. The observed $p_0$ value is \num{1.3E-5}.
The resulting significance is $\tZqsigMeas\sigma$, to be compared with the expected significance of $\tZqsigExpt\sigma$.
 
\section{Conclusion}
\label{sec:conclusion}
 
The cross-section for \tZq production has been measured using \lumi of proton--proton collision data collected
by the ATLAS experiment at the LHC in 2015 and 2016 at a centre-of-mass energy of $\sqrt{s} = \SI{13}{\TeV}$.
Evidence for the signal is obtained with a measured (expected) significance of $\tZqsigMeas\sigma$ ($\tZqsigExpt\sigma$).
The measured cross-section is \tZxsect.
This result is in agreement with the predicted SM \tZq cross-section, calculated at NLO to be \SI{800}{\fb}
with a scale uncertainty of $^{+6.1}_{-7.4}\%$.
 
\section*{Acknowledgements}
 
We thank CERN for the very successful operation of the LHC, as well as the
support staff from our institutions without whom ATLAS could not be
operated efficiently.
 
We acknowledge the support of ANPCyT, Argentina; YerPhI, Armenia; ARC, Australia; BMWFW and FWF, Austria; ANAS, Azerbaijan; SSTC, Belarus; CNPq and FAPESP, Brazil; NSERC, NRC and CFI, Canada; CERN; CONICYT, Chile; CAS, MOST and NSFC, China; COLCIENCIAS, Colombia; MSMT CR, MPO CR and VSC CR, Czech Republic; DNRF and DNSRC, Denmark; IN2P3-CNRS, CEA-DSM/IRFU, France; SRNSF, Georgia; BMBF, HGF, and MPG, Germany; GSRT, Greece; RGC, Hong Kong SAR, China; ISF, I-CORE and Benoziyo Center, Israel; INFN, Italy; MEXT and JSPS, Japan; CNRST, Morocco; NWO, Netherlands; RCN, Norway; MNiSW and NCN, Poland; FCT, Portugal; MNE/IFA, Romania; MES of Russia and NRC KI, Russian Federation; JINR; MESTD, Serbia; MSSR, Slovakia; ARRS and MIZ\v{S}, Slovenia; DST/NRF, South Africa; MINECO, Spain; SRC and Wallenberg Foundation, Sweden; SERI, SNSF and Cantons of Bern and Geneva, Switzerland; MOST, Taiwan; TAEK, Turkey; STFC, United Kingdom; DOE and NSF, United States of America. In addition, individual groups and members have received support from BCKDF, the Canada Council, CANARIE, CRC, Compute Canada, FQRNT, and the Ontario Innovation Trust, Canada; EPLANET, ERC, ERDF, FP7, Horizon 2020 and Marie Sk{\l}odowska-Curie Actions, European Union; Investissements d'Avenir Labex and Idex, ANR, R{\'e}gion Auvergne and Fondation Partager le Savoir, France; DFG and AvH Foundation, Germany; Herakleitos, Thales and Aristeia programmes co-financed by EU-ESF and the Greek NSRF; BSF, GIF and Minerva, Israel; BRF, Norway; CERCA Programme Generalitat de Catalunya, Generalitat Valenciana, Spain; the Royal Society and Leverhulme Trust, United Kingdom.
 
The crucial computing support from all WLCG partners is acknowledged gratefully, in particular from CERN, the ATLAS Tier-1 facilities at TRIUMF (Canada), NDGF (Denmark, Norway, Sweden), CC-IN2P3 (France), KIT/GridKA (Germany), INFN-CNAF (Italy), NL-T1 (Netherlands), PIC (Spain), ASGC (Taiwan), RAL (UK) and BNL (USA), the Tier-2 facilities worldwide and large non-WLCG resource providers. Major contributors of computing resources are listed in Ref.~\cite{ATL-GEN-PUB-2016-002}.
 
 
\printbibliography
 
\newpage
\begin{flushleft}
{\Large The ATLAS Collaboration}

\bigskip

M.~Aaboud$^\textrm{\scriptsize 137d}$,
G.~Aad$^\textrm{\scriptsize 88}$,
B.~Abbott$^\textrm{\scriptsize 115}$,
O.~Abdinov$^\textrm{\scriptsize 12}$$^{,*}$,
B.~Abeloos$^\textrm{\scriptsize 119}$,
S.H.~Abidi$^\textrm{\scriptsize 161}$,
O.S.~AbouZeid$^\textrm{\scriptsize 139}$,
N.L.~Abraham$^\textrm{\scriptsize 151}$,
H.~Abramowicz$^\textrm{\scriptsize 155}$,
H.~Abreu$^\textrm{\scriptsize 154}$
% ...
\bigskip
\\
% ...
$^{12}$ Institute of Physics, Azerbaijan Academy of Sciences, Baku, Azerbaijan\\
% ...
$^{88}$ CPPM, Aix-Marseille Universit{\'e} and CNRS/IN2P3, Marseille, France\\
% ...
$^{115}$ Homer L. Dodge Department of Physics and Astronomy, University of Oklahoma, Norman OK, United States of America\\
% ...
$^{119}$ LAL, Univ. Paris-Sud, CNRS/IN2P3, Universit{\'e} Paris-Saclay, Orsay, France\\
% ...
$^{137}$ $^{(a)}$ Facult{\'e} des Sciences Ain Chock, R{\'e}seau Universitaire de Physique des Hautes Energies - Universit{\'e} Hassan II, Casablanca; $^{(b)}$ Centre National de l'Energie des Sciences Techniques Nucleaires, Rabat; $^{(c)}$ Facult{\'e} des Sciences Semlalia, Universit{\'e} Cadi Ayyad, LPHEA-Marrakech; $^{(d)}$ Facult{\'e} des Sciences, Universit{\'e} Mohamed Premier and LPTPM, Oujda; $^{(e)}$ Facult{\'e} des sciences, Universit{\'e} Mohammed V, Rabat, Morocco\\
% ...
$^{139}$ Santa Cruz Institute for Particle Physics, University of California Santa Cruz, Santa Cruz CA, United States of America\\
% ...
$^{151}$ Department of Physics and Astronomy, University of Sussex, Brighton, United Kingdom\\
% ...
$^{154}$ Department of Physics, Technion: Israel Institute of Technology, Haifa, Israel\\
$^{155}$ Raymond and Beverly Sackler School of Physics and Astronomy, Tel Aviv University, Tel Aviv, Israel\\
% ...
$^{161}$ Department of Physics, University of Toronto, Toronto ON, Canada\\
% ...
$^{a}$ Also at Department of Physics, King's College London, London, United Kingdom\\
$^{b}$ Also at Institute of Physics, Azerbaijan Academy of Sciences, Baku, Azerbaijan\\
$^{c}$ Also at Novosibirsk State University, Novosibirsk, Russia\\
$^{d}$ Also at TRIUMF, Vancouver BC, Canada\\
$^{e}$ Also at Department of Physics {\&} Astronomy, University of Louisville, Louisville, KY, United States of America\\
% ...
$^{*}$ Deceased
\end{flushleft}

 
\end{document}